\documentclass[twocolumn,showpacs,preprintnumbers,amsmath,amssymb]{revtex4}
\usepackage{dcolumn}% Align table columns on decimal point
\usepackage{bm}% bold math
\usepackage{graphicx}

\begin{document}

\title{Astrophysical Measurement of the Equation of State of Neutron 
Star Matter}

\author{Feryal \"Ozel,$^1$ Gordon Baym,$^{2}$ and Tolga G\"uver$^1$}
\affiliation{$^1$University of Arizona, Department of Astronomy and 
Steward Observatory, 933 N. Cherry Ave., Tucson, AZ 85721\\
$^2$Department of Physics, University of Illinois, 1110 W. Green St.,
Urbana, IL 61801}

\begin{abstract} 
We present the first astrophysical measurement of the pressure of cold
matter above nuclear saturation density, based on recently determined
masses and radii of three neutron stars.  The pressure at higher
densities are below the predictions of equations of state that account
only for nucleonic degrees of freedom, and thus present a challenge to
the microscopic theory of neutron star matter.
\end{abstract}


\maketitle

Neutron stars probe the dense QCD phase diagram at lower temperatures
and higher baryon densities, in contrast to the higher
temperature--lower density regime in the early universe and in
ultrarelativistic heavy-ion collisions~\cite{qcd,jaipur}.  The baryon
density, $\rho$, in stellar interiors may reach an order of magnitude
beyond nuclear saturation density, $\rho_{\rm ns}\simeq 2.7\times
10^{14}$ g cm$^{-3} \simeq 0.16$ fm$^{-3}$, and cross into a regime
where quark degrees of freedom are excited or matter undergoes a meson
condensation phase transition.  The pressure of matter at these high
densities can, as we show here, be extracted from current neutron star
mass--radius determinations \cite{measmts}, and crucially constrains
calculations of high density neutron star matter.

The equation of state (EoS) of supranuclear matter determines the
dividing line between neutron stars and black holes, and directly
impacts the mechanism as well as outcomes of supernova explosions and
the numbers of neutron stars and black holes in the
Galaxy~\cite{supernova,KB}. In certain models, it affects the
mechanism and duration of gamma-ray bursts~\cite{collapsar}.  Accurate
evolutions of inspiraling neutron-star binaries and the collapse to
black holes, needed to calculate gravitational wave signals, depend
sensitively on the assumed EoS~\cite{gravwaves}.
   
Microscopic calculations of the EoS of neutron-star
matter have been based on a variety of inputs.  The approach most
firmly founded on experiment in the region of $\rho_{\rm ns}$ is to
determine two-body potentials from nucleon-nucleon scattering data
below 350 MeV and properties of light nuclei, supplemented by a
three-body potential~\cite{APR,mpr}.  Such calculations, accurate in
the neighborhood of $\rho_{\rm ns}$, have fundamental limitations.
Beyond a few times $\rho_{\rm ns}$ the forces between particles can no
longer be described via static few-body potentials; since the
characteristic range of the nuclear forces is $\sim 1/2m_\pi$, where
$m_\pi$ is the mass of the pion, the parameter measuring the relative
importance of three and higher body forces is $\sim\rho /(2m_\pi)^3
\sim 0.35\rho/\rho_{\rm ns}$.  Thus, at $\rho\gg\rho_{\rm ns}$ a well
defined expansion in terms of two-, three-, or more, body forces no
longer exists. EoS based on nucleons alone do not take into account
the rich variety of hadronic degrees of freedom that enter with
increasing density. In addition, pion condensates~\cite{APR,AM,PPT} or
kaon condensates~\cite{BB,GS} can enter at higher densities.
Field-theoretic models based on nucleons interacting via meson
exchange include, e.g., Ref~\cite{MS}; see Ref.~\cite{radius} for a
general summary of EoS. However, one cannot assume that matter at
higher densities can even be described in terms of well-defined
``asymptotic'' laboratory particles.  More realistically, one expects
in dense matter a gradual onset of quark degrees of freedom, not
accounted for by nucleons interacting via static potentials.  Indeed
once nucleons overlap considerably, the matter should percolate,
opening the possibility of their quark constituents propagating
throughout the system.  Such additional degrees of freedom should lead
to softening of the EoS, consistent with our findings here, and thus
to a lower maximum neutron star mass.  Owing to difficulties of
determining the neutron star matter EoS from first principles, neutron
star observations become an important input in determining the EoS at
high density, and constraining microscopic calculations.

A number of astrophysical observations have focused on measuring neutron
star radii, $R$, and masses, $M$, in an attempt to constrain the
uncertainties in the  EoS (summarized in Ref.~\cite{radius}). 
The measurement of post-Newtonian parameters of double
neutron stars provide precise determinations of their
masses~\cite{pulsars}. Glitches observed in radio pulsars lead to
constraints on the moment of inertia, and therefore, on neutron star
masses and radii~\cite{glitches}. Observations of the thermal emission
from accreting neutron stars in quiescence and from millisecond X-ray
pulsars result in broad, correlated constraints on neutron star masses
and radii~\cite{obs1}.

Recent advances in both astrophysical techniques and neutron-star
atmosphere modeling allow us for the first time to measure the
pressure of neutron star matter at supranuclear densities directly
from observations. References~\cite{radius} suggested that the radii of
neutron stars are a good indicator of the pressure at roughly twice
the nuclear saturation density. This argument was further extended in
\cite{inversion1,inversion2}, which demonstrated that knowing
the properties of ultradense matter at three fiducial densities allows
one to reproduce macroscopic neutron star properties, including the
mass-radius relation and stellar moment of inertia.  Conversely, three
distinct measurements of neutron star masses and radii, as we use
here, is sufficient to infer a {\it piecewise polytropic} EoS of
matter at supranuclear densities~\cite{inversion1}.

Observations of multiple spectroscopic phenomena during thermonuclear
bursts from X-ray binaries have enabled the tightest measurements
neutron star radii and masses to date~\cite{measmts}. The long-term
monitoring of burst sources with the Rossi X-ray Timing Explorer, with
its excellent photon statistics, has resulted in a large ($> 1000$)
database of bursts~\cite{catalog}, from which systematic uncertainties
can be determined and controlled, and ideal sources that act as
standard candles can be identified. High resolution X-ray spectroscopy
with the Chandra X-ray Observatory and XMM-Newton has led to a
detailed measurement of the soft X-ray spectra of bursters and reduced
the uncertainties introduced by interstellar extinction~\cite{ism}.
Finally, pointed optical/infrared observations with the Hubble Space
Telescope and large ground-based facilities (e.g., the Magellan
telescope) have substantially improved distance measurements to these
sources (see Ref.~\cite{measmts} and references therein).

The wealth of such high quality data allows us to employ a novel
approach, combining different spectroscopic measurements to break the
degeneracies between neutron star masses and radii inherent to each
observable~\cite{method1,method2}. The first observable is the
apparent surface area during the cooling phase of the bursts,
\begin{equation}
A=\frac{R^2}{D^2f_{\rm c}^4}\left(1-\frac{2GM}{Rc^2}\right)^{-1}\;,
\label{A}
\end{equation}
where $D$ is the
distance to the source, and $f_{\rm c}$ is a calculated ratio between
the spectral (color) and effective temperature, $T$, of the emerging
radiation that accounts for the non-Planckian spectrum of the
burst. Because the emitted luminosity is $\propto T^4$, the apparent
surface area in Eq.~(\ref{A}) shows the same $T$ dependence, which we
absorb into the definition of $f_{\rm c}$. The apparent surface area
remains constant in time and is highly reproducible in multiple events
from the same source, indicating that the entire neutron star surface,
rather than a variable area on the surface, participates in the burst
emission.

The second phenomenon occurs in a subset of bursts, when the flux
becomes so high that it exceeds the local Eddington limit and lifts
the photosphere of the neutron star.  The flux achieved during these
events is also highly reproducible for a large number of sources
including the three discussed below, and is related to the neutron
star mass and radius through
\begin{equation}
F_{\rm Edd}=
\frac{GMc}{k_{\rm es}D^2}\left(1-\frac{2GM}{Rc^2}\right)^{1/2}. 
\end{equation}
This ``touchdown" flux is evaluated at the moment when the photosphere
has receded back to the neutron star surface (see also
Ref.~\cite{measmts}).  In the above equation, $k_{\rm es}=0.2
(1+X)$~cm$^2$~g$^{-1}$ is the electron scattering opacity in the
stellar atmosphere. In the atmospheric models, we considered a wide
range of hydrogen mass fraction $X$ consistent with the properties of
each binary system.

\begin{figure}
\centering
   \includegraphics[scale=0.45]{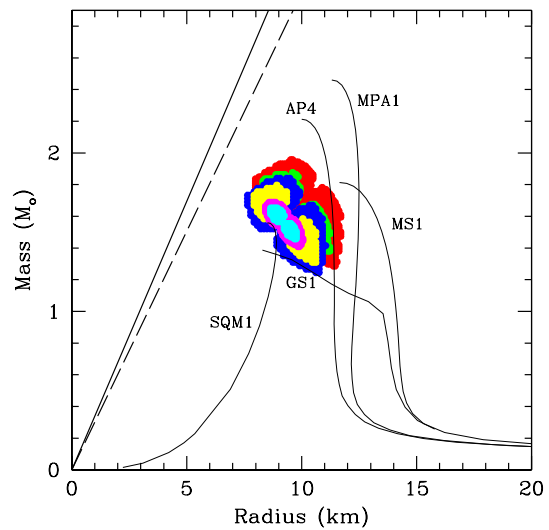}
\caption{The 1- and 2-$\sigma$ confidence contours for the masses and 
radii of three neutron stars in the binaries 4U\,1608$-$248
(green/red), EXO~1745$-$248 (yellow/blue), and 4U\,1820$-$30
(cyan/magenta), compared with predictions of representative EoS (see
text for details). The details of the measurements are described in
Refs.~\cite{measmts}. The diagonal lines are the black-hole event
horizon (solid) and Buchdahl (dashed) \cite{buchdahl} limits.}
\label{fig:mr} 
\end{figure}

Combining the measurements of $A$ and $F$ with the distance $D$ to
each source, we obtain tight, uncorrelated constraints on the masses
and radii of neutron stars. We have applied this technique to three
sources, the neutron stars in the binaries 4U\,1608$-$248,
EXO~1745$-$248, and 4U\,1820$-$30, and show in Fig.~\ref{fig:mr} the
1- and 2-$\sigma$ confidence contours of their masses and radii
determinations~\cite{measmts}. The results are a set of uncorrelated
measurements of neutron star masses and radii. (An earlier measurement
of the mass and radius of the neutron star in
EXO~0748$-$676~\cite{method1} was based on the identification of
atomic line features in its X-ray spectrum with gravitationally
redshifted lines from its surface, which has since been shown to be
inconsistent~\cite{exo0748_2} with the recent measurement of its rapid
spin frequency~\cite{gall10}).

The measurements in Fig.~\ref{fig:mr} incorporate the corrections and
systematic uncertainties associated with the modeling of emission from
the hot surfaces of neutron stars following a thermonuclear burst, the
composition of the neutron star surface, as well as statistical or
systematic uncertainties in the distances to the binaries. In all
three measurements, uncertainties arising from subtraction of the
background flux are negligible, because in each source, the luminosity
from the neutron star surface exceeds the accretion luminosity by more
than a factor of ten. Since not every combination of observables leads
to a solution for $M$ and $R$, we converted the probability densities
over the measured fluxes, apparent areas, and distances to those over
the neutron star mass and radius following standard Bayesian
statistics~\cite{measmts}.  As a result, the uncertainties in the mass
and radius, of order 15\%, are smaller than those of the individual
spectroscopic quantities. Note that the confidence contours shown in
Fig.~\ref{fig:mr} correct a small numerical error in the Jacobian
transformation of Refs.~\cite{measmts}.

\begin{figure}[t]
\centering
   \includegraphics[scale=0.45]{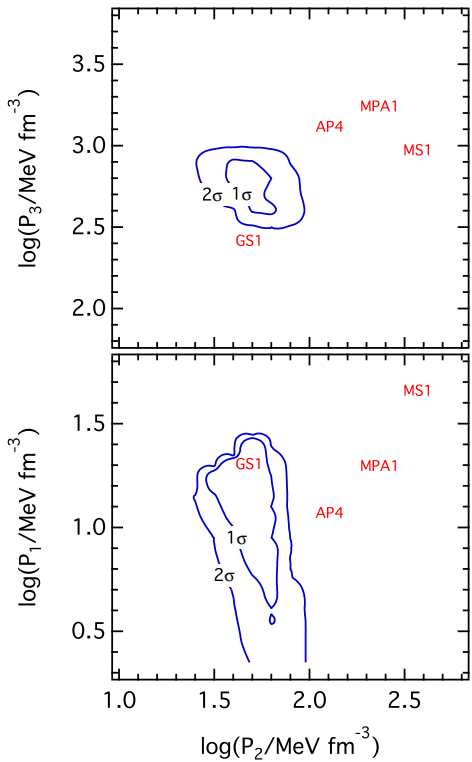}
\caption{The pressure of cold matter at {\em (top)\/} 7.4 and 
3.7 $\rho_{\rm ns}$ and {\em (bottom)\/} 1.85 and 3.7 $\rho_{\rm ns}$.}
\label{p1p2} 
\end{figure}

As noted, three distinct measurements of neutron star masses and radii
allow us to infer a piecewise EoS of matter at supranuclear
densities~\cite{inversion1}. This approach makes the explicit
assumption that the density in the neutron star surface layers
smoothly reduces to zero, and is, therefore, not applicable to strange
quark matter stars that are not gravitationally bound. In particular,
we follow the procedure in Ref.~\cite{inversion1} to convert these
measurements to probability densities over the pressures of neutron
star matter at three fiducial baryon densities $\rho_1 = 1.85
\rho_{\rm ns}$, $\rho_2 = 3.7 \rho_{\rm ns}$, and $\rho_3 = 7.4
\rho_{\rm ns}$.  In these calculations, we include the full
probability density for all three sources. We supplement this
procedure with the requirement of causality, rejecting combinations of
pressures for which the sound speed is larger than the speed of
light~\cite{inversion1}. Fig.~\ref{p1p2} shows the confidence
contours of different pairs of pressures (integrated over the third
pressure).

The pressure at $\rho_1$ is only weakly constrained because such low
densities are important only in determining the macroscopic properties
of neutron stars with masses smaller than those in our measured
sample~\cite{inversion1}.  In contrast, the pressures at $\rho_2$ and
$\rho_3$ are constrained to within a factor of $\sim 10^{0.3}$.

In detail, we describe these data with a phenomenological piecewise
polytropic EoS above density $\rho_0 = 10^{14}$~g~cm$^{-3}$, which we
smoothly connect to the SLy EoS at lower densities~\cite{Douchin}.
Between $\rho_0$ and $\rho_4$ we fit the energy density in the
interval $\rho_{i-1}<\rho\leq\rho_i$ as $\epsilon = \alpha_i\rho +
\beta_i \rho^{\Gamma_i}, $ from which we derive the pressure, $P =
\rho^2\partial(\epsilon/\rho)/\partial \rho = (\Gamma_i-1) \beta_i
\rho^{\Gamma_i}$.  Since the Tolman-Oppenheimer-Volkoff equation 
relates $\epsilon$ and $P$ to the mass and radius of the star, one 
only can determine the baryon density $\rho$ from $\epsilon$ and $P$ 
data to within a scale factor; to determine the scale we connect our 
fit to the tabulated low density SLy EoS. For $\rho_{i-1}<\rho\leq\rho_i$, 
the effective polytropic index is $
\Gamma_i \equiv \log(P_i/P_{i-1})/\log(\rho_i/\rho_{i-1})$, 
the pressure is
\begin{equation}
P = P_i \left(\frac{\rho}{\rho_i}\right)^{\Gamma_i}, 
\label{pgamma}
\end{equation}
and in the energy density 
%\begin{equation}
%\epsilon = (1+a_i)\rho + \frac{P_i}{\Gamma_i-1} \left(\frac{\rho}{\rho_i}
%\right)^{\Gamma_i}, 
%\end{equation}
%where 
\begin{equation}
\alpha_i = \frac{\epsilon_{i-1}}{\rho_{i-1}}-\frac{P_i}{(\Gamma_i-1)
\rho_{i-1}} \left(\frac{\rho_{i-1}}{\rho_i}\right)^{\Gamma_i}. 
%a_i = \frac{\epsilon_{i-1}}{\rho_{i-1}}-1-\frac{P_i}{(\Gamma_i-1)
%\rho_{i-1}} \left(\frac{\rho_{i-1}}{\rho_i}\right)^{\Gamma_i}. 
\end{equation}
Table~1 shows the fitting parameters. For $\rho > \rho_3$, we
extrapolate the last polytropic relation.

\begin{table}[t]
\caption{\label{tab:table1} }
Measured pressures at three supranuclear densities, in MeV~fm$^{-3}$,
together with $P_0$ taken from the low density SLy EoS~\cite{Douchin}.
\newline
\begin{ruledtabular}
\begin{tabular}{cccc}
log $P_0$($0.37 \rho_{\rm ns}$) & 
log $P_1$($1.85 \rho_{\rm ns}$) &  
log $P_2$($3.7 \rho_{\rm ns}$) & 
log $P_3$($7.4 \rho_{\rm ns}$) \\
\hline 
-0.64 & [0.6--1.4] & $1.70^{+0.15}_{-0.15}$ & $2.8^{+0.1}_{-0.2}$ \\
\end{tabular}
\end{ruledtabular}
\label{table: EoS}
\end{table}

\begin{figure}[t]
\centering
   \includegraphics[scale=0.45]{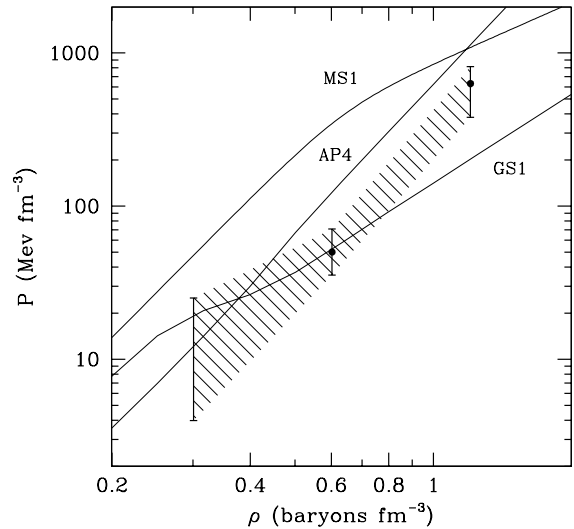}
\caption{Pressure vs. baryon density at the three points, 
$P_1$, $P_2$, and $P_3$, together with the fitted  EoS
(\ref{pgamma}). The shaded region shows the uncertainties in the
determination. }
\label{pressures} 
\end{figure}

The data present a clear challenge to microscopic nuclear
calculations.  Figs.~\ref{p1p2} and \ref{pressures} compare the best
fit values of the pressures at the three fiducial densities with those
predicted by a representative sample of EoS based on a wide range of
input physics and computational methods, from nucleonic: variational
chain summation with the AV18 potential, UIX three-body potential plus
relativistic boost corrections {\it AP4} \cite{APR};
Dirac-Brueckner-Hartree-Fock {\it MPA1} \cite{radius}; and
relativistic mean fields {\it MS1} \cite{MS}, plus kaons {\it GS1}
\cite{GS}. (Although our procedure is not applicable 
to the u,d,s quark matter {\it SQM1} \cite{radius}, its predictions
can nevertheless be compared directly to the individual mass and
radius measurements shown in Fig.~\ref{fig:mr}).  Our measurements
clearly discriminate between different predictions, and indicate that
the EoS based on nucleons alone, AP4, MP1 and MS1, are too stiff at
higher density -- a conclusion also borne out by the comparisons of
$M$ vs.~$R$ in Fig.~\ref{fig:mr} which show that the predicted radii
at the measured masses are too large.  As one sees clearly in
Fig.~\ref{pressures}, the data call for a softer EoS, as would be
produced by including degrees of freedom beyond nucleons, e.g.,
hyperons, mesons, and quarks, or possibly produced by a better
description of nucleonic interactions.  In particular a softer EoS, by
allowing larger central densities than an EoS with only two and three
body nucleonic interactions (e.g., AP4 with a central density $\sim
7\rho_{ns}$ at the maximum mass $\sim 2.2 M_\odot$) would, in the
naive picture of a sharp phase transition from nucleonic to quark
matter, allow quark cores to appear in more massive stars.

The allowed range of pressures seen in Fig.~\ref{pressures} also
reduces the maximum neutron star mass compared to that predicted by
purely nucleonic EoS. It is, nevertheless, consistent with the recent
measurement of a $1.97\pm0.04~M_\odot$ neutron star by observations of
Shapiro delay \cite{psr}.

Our measurements can be confirmed or tested by additional mass-radius
measurements in bursting sources, improved distance measurements by
space-based interferometers, and more definitively by observations of
other phenomena that probe the masses and radii of neutron stars such
as gravitationally redshifted absorption lines or flux oscillations
that depend on surface gravity.

\begin{center}
{\bf Acknowledgements}
\end{center}
F. \"O. thanks the Institute for Theory and Computation at Harvard
University for their hospitality. We gratefully acknowledge useful
conversations with A. Loeb, R. Narayan, D. Psaltis and K. Rajagopal.
F. \"O. acknowledges support from NSF grant AST-07-08640 and Chandra
Theory grant TMO-11003X. This research was supported in part by NSF
Grant PHY-07-01611.

\end{document}